# Surface acoustic wave modulation of single photon emission from GaN/InGaN nanowire quantum dots

S Lazić[1], E Chernysheva[1], A Hernández-Mínguez[2], P V Santos[2] and H P van der Meulen[1]

[1] Departamento de Física de Materiales, Instituto 'Nicolás Cabrera' and Instituto de Física de Materia Condensada (IFIMAC), Universidad Autónoma de Madrid, 28049 Madrid, Spain
[2] Paul-Drude-Institut für Festkörperelektronik, Hausvogteiplatz 5-7, 10117 Berlin, Germany

lazic.snezana@uam.es

**Abstract**. On-chip quantum information processing requires controllable quantum light sources that can be operated on-demand at high-speeds and with the possibility of in-situ control of the photon emission wavelength and its optical polarization properties. Here, we report on the dynamic control of the optical emission from core-shell GaN/InGaN nanowire (NW) heterostructures using radio frequency surface acoustic waves (SAWs). The SAWs are excited on the surface of a piezoelectric lithium niobate crystal equipped with a SAW delay line onto which the NWs were mechanically transferred. Luminescent quantum dot (QD)-like exciton localization centers induced by compositional fluctuations within the InGaN nanoshell were identified using stroboscopic micro-photoluminescence (micro-PL) spectroscopy. They exhibit narrow and almost fully linearly polarized emission lines in the micro-PL spectra and a pronounced anti-bunching signature of single photon emission in the photon correlation experiments. When the nanowire is perturbed by the propagating SAW, the embedded QD is periodically strained and its excitonic transitions are modulated by the acousto-mechanical coupling, giving rise to a spectral fine-tuning within a ~1.5 meV bandwidth at the acoustic frequency of ~330 MHz. This outcome can be further combined with spectral detection filtering for temporal control of the emitted photons. The effect of the SAW piezoelectric field on the QD charge population and on the optical polarization degree is also observed. The advantage of the acousto-optoelectric over other control schemes is that it allows in-situ manipulation of the optical emission properties over a wide frequency range (up to GHz frequencies).

## 1. Introduction

The implementation of semiconductor heterostructures on a nanowire (NW) platform provides scalable miniaturized architecture for nanoscale photonic and optoelectronic devices and facilitates their integration into modern on-chip technologies. However, to fully exploit the potential of NW-based devices, appropriate techniques are required to allow in situ control and manipulation of their fundamental physical properties on a nanoscale. Compared to other control schemes, which typically require selective doping and sophisticated nanofabrication of electrical contacts on individual NWs with sub-micrometer dimensions, the acousto-optoelectric approach based on radio-frequency (rf) surface acoustic waves (SAWs) has emerged as a powerful contactless tool for reliable manipulation of the optical emission characteristics from small-scale NW heterostructures, including NW quantum dots (QDs). Moreover, because the SAWs propagate at the speed of sound, high operating frequencies (from several tens of megahertz up to several tens of gigahertz) can be achieved, which are usually inaccessible by other control schemes.

Most reports on SAW-modulated NW QD structures to date are limited to III-arsenide material systems. In this contribution, we review our experiments on the effect of SAW-induced strain and piezoelectric fields on the photon emission dynamics from wurtzite III-nitride dot-in-a-nanowire heterostructures. Applying SAW-techniques to group III-nitrides is vital, as their wide bandgap, large exciton binding energy and band offsets make them an ideal candidate for high-power and high-temperature device applications.

## 2. Methods

A schematics of the sample configuration is displayed in Fig. 1(a). All optical experiments are performed on a 128° Y-cut lithium niobate (LiNbO$_3$) substrate photo-lithographically patterned with an acoustic delay line consisting of two interdigital transducers (IDTs). The IDTs are designed to generate SAWs with an acoustic wavelength $\lambda_{SAW}$=11.67 µm, corresponding to a room-temperature resonant frequency of $f_{SAW}$~332 MHz and temporal period of $T_{SAW}$~3 ns. Information on the rf reflection and transmission characteristics of the SAW-chip can be found in [1]. The epitaxially grown semiconductor NWs containing a pencil-like GaN core, an InGaN nanoshell and an external conformal GaN cap were mechanically transferred onto the area between the two IDTs. The intrawire InGaN nanoshell hosts QD-like exciton localization centers are induced by facet-dependent indium incorporation and composition fluctuation. More details on the NW growth and QD formation mechanism are given elsewhere [2,3].

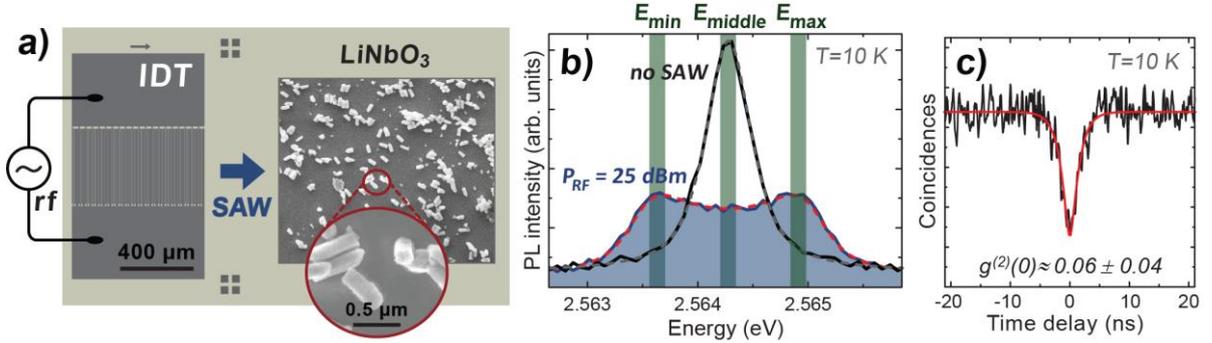

**Figure 1.** (a) Layout of the SAW-chip formed by IDTs patterned on the surface of a LiNbO$_3$ crystal containing GaN/InGaN NWs dispersed on the SAW propagation path. Inset: Magnified view of the optically probed NWs. (b) Time-integrated micro-PL spectrum of a single NW QD without (unshaded trace) and with (shaded trace) SAW applied. The rectangles labeled $E_{min}$, $E_{middle}$ and $E_{max}$ mark the spectral detection windows used in Fig. 2(a) (see text). (c) Autocorrelation histogram of the unperturbed (i.e. no SAW) emission peak in (b).

The effect of the SAW on the optical emission from NW QDs formed on the topmost polar *c*-facet of the InGaN nanoshell was characterized by low-temperature (T~10 K) spatially, time- and polarization-resolved micro-PL spectroscopy. The experiments were performed under direct optical excitation of single dispersed NWs. To monitor the acousto-mechanic and acousto-electric effects on the probed NW QD emission dynamics we employ a stroboscopic micro-PL technique with either time-integrated or time-resolved detection [1,4]. The photon emission statistics was analyzed using a Hanbury Brown and Twiss (HBT) setup, which delivers a *histogram of the photon coincidences* as a function of the arrival time delay at the two HBT detectors [1]. When triggered by the laser pulses, the same setup also yields time-resolved PL traces.

## 3. Results and discussions

Figure 1(b) presents low-temperature micro-PL spectrum of one of the optically excited NW QDs exhibiting strong electromechanically coupling to the propagating SAW [5]. In the absence of the SAW, the spectrum (black trace) recorded under continuous wave (cw) optical excitation at 442 nm shows a sharp emission line due to the ground-state exciton (X) transition [1]. Its autocorrelation histogram in Fig. 1(c) reveals a pronounced signature of single photon emission: an anti-bunching dip at zero-time delay ($\tau$=0) with the $g^{(2)}(0)$ value of 0.06±0.04 [1]. When the SAW is applied, it couples to the probed NW QD through the deformation potential mechanism and, consequently, periodically modulates the X's emission energy [4]. Averaged over time, this modulation gives rise to an apparent splitting of the X emission line (cf. Fig. 1(b), red trace). Due to the time-integrated detection and the unlocked excitation scheme (i.e. cw excitation), the recorded data provides an averaged picture of the SAW-governed carrier recombination. In this way, we can assess the full bandwidth of the SAW-driven periodic spectral tuning of the QD optical transitions, which for the highest SAW intensity at $P_{rf} =+25$ dBm reaches a maximum value of $\Delta E =1.55$ meV [4,5].

By collecting the photons in a narrow energy window (cf. Fig. 2(a)), the SAW-driven spectral tuning forces the X emission energy to move in and out of the selected detection range and, consequently, changes in the photon count rate are detected at one of the HBT detectors within each SAW cycle. As seen in Fig. 2(a), light collected at the highest ($E_{max}$) and lowest ($E_{min}$) energy of the SAW-induced dynamic spectral modulation is of opposite phase and is emitted at the SAW frequency. In contrast, at the central wavelength ($E_{middle}$), the emission occurs at twice the acoustic frequency, as the X peak moves past this energy twice per acoustic cycle. This spectral detection filtering enables the photon output to be clocked at the acoustic frequencies and can thus be employed to control the photon emission time. Such acoustically regulated control of the photon emission time provides an efficient mechanism for deterministic on demand generation of single photons without the need for a pulsed laser. By increasing the SAW frequency, high repetition rate of the emitted photons can be achieved.

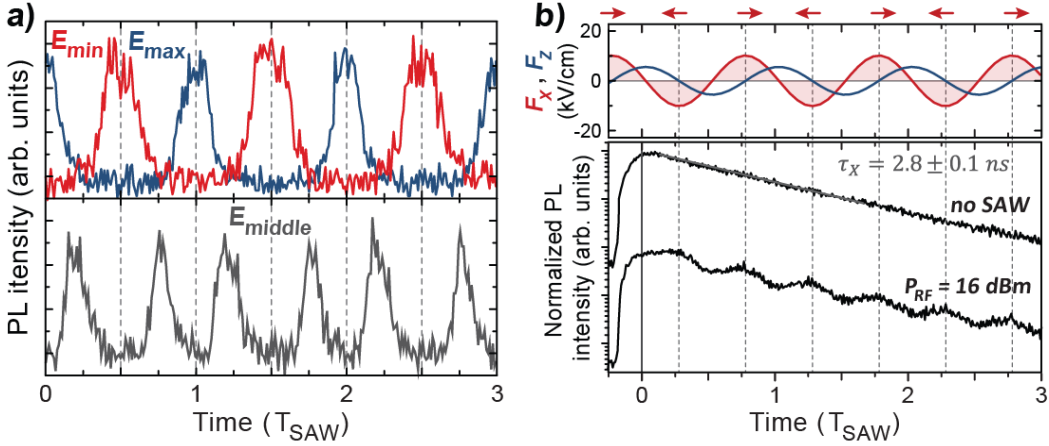

**Figure 2.** (a) Time-resolved PL intensity of the probed NW QD emission recorded under stroboscopic detection at spectral positions marked by the corresponding rectangles labeled $E_{min}$, $E_{middle}$ and $E_{max}$ in Fig. 1(b). (b) Lower panels: PL decay curves of the NW QD emission with no SAW applied (upper trace, showing an exciton lifetime of $\tau_X =2.8\pm0.1$ ns) and under SAW excitation with $P_{rf} =+16$ dBm (lower trace). The traces are vertically offset for clarity. Upper panels: time dependence of the SAW longitudinal $F_X$ (red shaded curves) and transverse $F_Z$ (blue curves) electric field components at the position of the probed NW QD.

To better understand the underlying dynamics of the excitonic transition in the acoustically modulated QD confinement potential, we registered the time evolution of its PL under SAW-synchronized pulsed optical excitation and time- and spectrally resolved detection conditions. The lower panel in Fig. 2(b) presents the temporal decay at a moderate acoustic power of the PL, collected over the entire spectral modulation bandwidth of the SAW-split X line in Fig. 1(b). The measured time-resolved PL profiles reveal the dependence of the SAW-mediated QD emission dynamics as a function of the local phase of the SAW. Contrary to the unperturbed PL transients (upper trace in lower panel of Fig. 2(b)), the stroboscopic time-resolved measurement shows weak oscillations in the PL emission intensity during the decay. For a sufficiently slow decay of the order of the acoustic period (i.e. $\tau_X \approx T_{SAW}$), these PL intensity oscillations appear twice per SAW cycle, yielding multiple characteristic signatures in the X PL time-transient. These increased PL intensity signatures coincide with the acoustic phases marked with vertical dashed lines and horizontal red arrows in Fig. 2(b). As seen from the numerical modelling (see Ref. [4] for more details) of the temporal evolution of the longitudinal $F_X$ (i.e. along SAW propagation direction) and transverse $F_Z$ (i.e. perpendicular to the substrate surface) components of the SAW-induced electric field for $P_{rf} =+16$ dBm, these marked SAW phases correspond to the maximum and minimum values of $F_X$. Considering that the studied QD is, essentially, a potential trap induced by indium content fluctuations within the topmost polar InGaN nanoshell section, it acts as an optically active deep trap for electrons and holes photo-excited in the InGaN region. We, thus, relate the observed PL intensity oscillations to the photo-generation in a continuum of states of the InGaN nanoshell region surrounding the QD, where the SAW's $F_X$ induces, at phases corresponding to its maximum magnitude, simultaneous transfer of photo-excited electrons and holes into the energetically lower QD states [4]. The transverse electric field component $F_Z$, on the other hand, has negligible effect on the charge population of the SAW-modulated QD potential owing to its small contribution compared to the strong built-in electric field acting along the axis of wurtzite III-nitride NWs [5].

The effect of the SAW piezoelectric field on the polarization degree of the QD optical emission is also observed and is presented by the substantial reduction (up to 30%, not shown) of the initially high linear polarization degree (~80%) with increasing the SAW intensity. For our NW QDs, previous measurements on the same batch as the one studied here have shown that the SAW fields can also tune the biexciton binding energy [5]. As the biexciton binding energy is determined by electron-hole Coulomb interactions in the QD, this shows that the SAW fields can as well affect various aspects of the QD band structure, such as band energy separation or mixing. We, hence, attribute the observed behavior to the SAW-induced change in the QD valence band structure (unpublished results). In fact, a significant change in the linear polarization degree has been predicted in Ref. [6] as a function of the ratio between the in-plane QD anisotropy parameter ($\Xi_{6a}$) and the in-plane component of the spin-orbit interaction ($\Delta_2$). Actually, in wurtzite III-nitride QDs, as a result of the small energy separation between the two uppermost valence bands (i.e. small $\Delta_2$), even a small in-plane anisotropy is able to produce large values of the polarization degree. In addition, it has been shown that the spin-orbit coupling can be effectively manipulated in InAs QDs by an external electrical bias [7]. Thus, for the underlying mechanism of the observed reduction in the photon polarization degree we propose the modification of the spin-orbit coupling by the SAW piezoelectric field.

In summary, we have demonstrated the SAW-induced fine-tuning of the III-nitride NW QDs emission energy and combined it with the spectral detection filtering to control the photon output timing. We have also shown that the SAW piezoelectric field can be employed to control the NW QD charge population and its optical polarization degree. The latter is particularly relevant for quantum communication and information processing applications for which real-time control of the photon polarization state is required.

**Acknowledgements**

The authors thank J.M. Calleja form the Universidad Autónoma de Madrid for scientific discussions & Ž. Gačević and E. Calleja from ISOM-DIE at the Universidad Politécnica de Madrid for the epitaxial growth of nanowire heterostructures. E.C. (S.L.) acknowledges the Spanish MINECO FPI (RyC-2011-09528) grant. The project was partially funded by the Spanish MINECO Grants MAT2014-53119-C2-1-R and MAT2017-83722-R.